\documentclass[11pt,twoside]{article}
\usepackage{asp2006}
\usepackage{epsf}
\usepackage{graphics}
\usepackage{lscape}
\usepackage{natbib}

\begin{document}

\title{Optical and X-ray Variability of AGNs}

\author{C. Martin Gaskell}
\affil{Department of Physics \& Astronomy, University of Nebraska,
Lincoln, NE 68588-0111, USA}

\begin{abstract}
I present new comparisons of AGN optical, UV, and X-ray variations.
These reveal complex relationships between the different passbands
that can change with time in a given object.  While there is
evidence from several objects that X-ray and optical activity levels
are correlated on long timescales, variations on shorter timescales
can occur independently.  It is proposed that the combination of
correlated and uncorrelated short-timescale variability is a
consequence of anisotropic high-energy emission.  It is also argued
that the correlation between X-ray and optical variability on long
timescales must be due to a common underlying factor and not to
reprocessing of X-ray radiation.
\end{abstract}

\section{Introduction}

\citet{Lyuty78} showed that the X-ray and optical continuum fluxes
of NGC~415 tracked one another on a timescale of months. The general
expectation until fairly recently has been that the X-ray and
optical would correlate with each other \citep[e.g.,][]{Collin91}.
However, in recent years there have been various cases where the
X-rays have {\it not} correlated with the optical.  The most
interesting of these has been NGC~3516 \citep{Maoz02}. Here I give
some results of our optical/X-ray monitoring of three very different
AGNs and propose a mechanism to explain the curious combination of
correlated and uncorrelated optical and X-ray variability seen in
AGNs.

\subsection{NGC 5548 -- a Typical Broad-Line Seyfert 1}

In 1998 we carried out a short {\it RXTE}/{\it ASCA}/{\it
EUVE}/optical observing campaign \citep{Chiang00,Dietrich01}. During
this campaign the {\it EUVE} flux led all other passbands. The {\it
RXTE} PCA 2--10 keV flux lagged the {\it EUVE} by $0.39 \pm 0.05$
days. The {\it ASCA} SIS 0.5--1 keV flux lagged the {\it EUVE} by
$0.15 \pm 0.03$ days. Combining the delays for all optical passbands
in \citet{Dietrich01} we find the optical lagging the {\it EUVE} by
$0.2 \pm 0.2$ days. However, while the optical and {\it EUVE} show
the same trend over our 30 day monitoring period, during our most
intensive 5 days of simultaneous monitoring the detailed agreement
is poor.  The correlated flux variability between the X-ray and EUV
fluxes in 1998 could be merely fortuitous. \citet{Haba03} give an
example of where there is a factor of two dip in the EUV flux over a
few hours while the {\it ASCA} X-ray flux remains constant.

\citet[][see also these proceedings]{Uttley03} have shown that there
is a good correlation on a long ($>$ one month) timescale between
the hard (2--10 keV) X-ray flux and the optical flux, as had been
found for NGC~4151 by \citet{Lyuty78}. \citet{Peterson00} found that
the optical and X-ray fluxes of NGC 4051 were also correlated on
similar long timescales. Even for NGC~3516, there is a correlation
of the X-ray and optical fluxes from year to year. Interestingly,
for NGC 5548, Uttley et al. find that the long-term optical
amplitude is slightly {\it greater} than the X-ray amplitude.

In July of 2001 we carried out intensive monitoring of NGC~5548
(Gaskell, Klimek et al. in preparation) in conjunction with
intensive {\it RXTE} monitoring.  Observational details are as in
\citet{Klimek04}. In Fig. 1 we show the host-galaxy-subtracted
Nebraska V-band photometry compared with the {\it RXTE} 2--10 keV
fluxes.  It can be seen that while the general trend of the optical
variations (shown by the smooth curve) agrees with the long-term
trend of the X-rays (and, interestingly, with no detectable lag),
there are major short-term X-ray events (a ``flare'' and an
``anti-flare'') that are {\it completely absent} in the optical
light curve.

\begin{figure}[t!]
\begin{center}
\epsfxsize = 120 mm \epsfbox{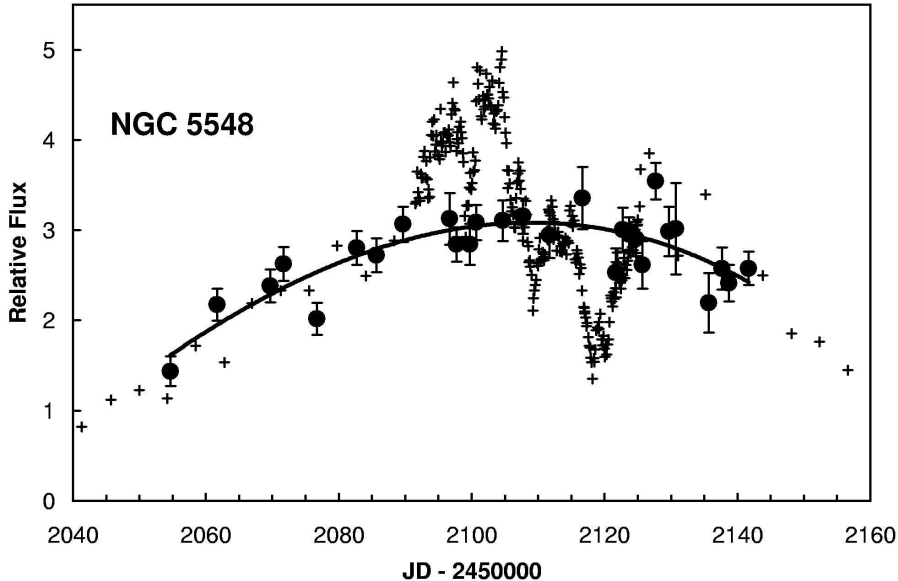} \caption{A
comparison of the optical and X-ray variability of NGC~5548 over a
three-month period in 2001.  The {\it RXTE} 2--10 keV flux is shown
by small crosses, and University of Nebraska V-band photometry
(corrected for the host galaxy flux) is given by the solid circles
with error bars. \label{NGC5548}}
\end{center}
\end{figure}

\subsection{The NLS1 Ark 564}

Ark 564 is our best-studied Narrow-Line Seyfert 1 (NLS1).  Fig. 2
shows the results of our simultaneous {\it ASCA}, {\it HST}, and
optical monitoring of the narrow-line Seyfert 1 galaxy Ark 564. The
1 keV {\it ASCA} data are a 0.3-day running average taken from the
observations reported in \citet{Turner01} and \citet{Edelson02}. The
$\lambda$1365 UV fluxes are from \citet{Collier01}, and the optical
photometry from \citet{Shemmer01}. It can be seen that there are
numerous X-ray events on sub-day timescales and some of the largest
of these have lower-amplitude UV and optical events following them
by $\sim 1.5$ days.

\begin{figure}[t!]
\begin{center}
\epsfxsize = 120 mm \epsfbox{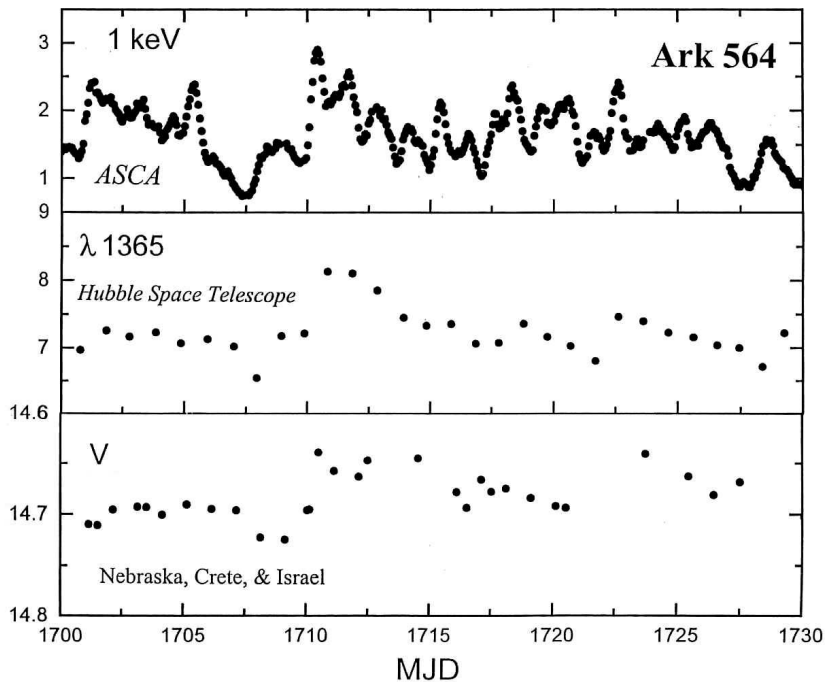} \caption{X-ray, UV,
and optical variability of Ark 564.  The top panel shows 0.3-day
running average 1 keV {\it ASCA} fluxes, the middle panel {\it HST}
$\lambda1365$ UV fluxes, and the bottom panel V-band magnitudes
(with no host galaxy correction) from the University of Nebraska,
Skinakas Observatory (Crete), and Wise Observatory (Israel).
\label{Ark564}}
\end{center}
\end{figure}

There are several interesting points to note.  First, the UV and
optical track each other well. Second, the UV and optical
variability was of much lower amplitude than the X-ray variability.
(After host galaxy correction, the optical and UV amplitudes are
probably similar.)  Third, as with NGC 5548, there were X-ray events
with no optical counterparts (notably the short flare at JD
2451706).

\subsection{The Broad-Line Radio Galaxy 3C 390.3}

3C 390.3 is our best-studied radio galaxy.  Radio galaxies lie at
the opposite end of the eigenvector 1 from NLS1s.  Fig. 3 shows the
results of our simultaneous soft X-ray, UV, and optical monitoring.
The {\it ROSAT} 0.1 -- 2 keV fluxes are from \citet{Leighly97}, the
{\it IUE} $\lambda$1370 UV fluxes are based on \citet{OBrien98}, and
V-band fluxes have been derived with scalings and corrections from
\citet{Dietrich98}.  3C 390.3 has a substantial (and as yet
undetermined) host galaxy contribution. To emphasize the optical
variations we have subtracted an arbitrary constant flux.

\begin{figure}[t!]
\begin{center}
\epsfxsize = 120 mm \epsfbox{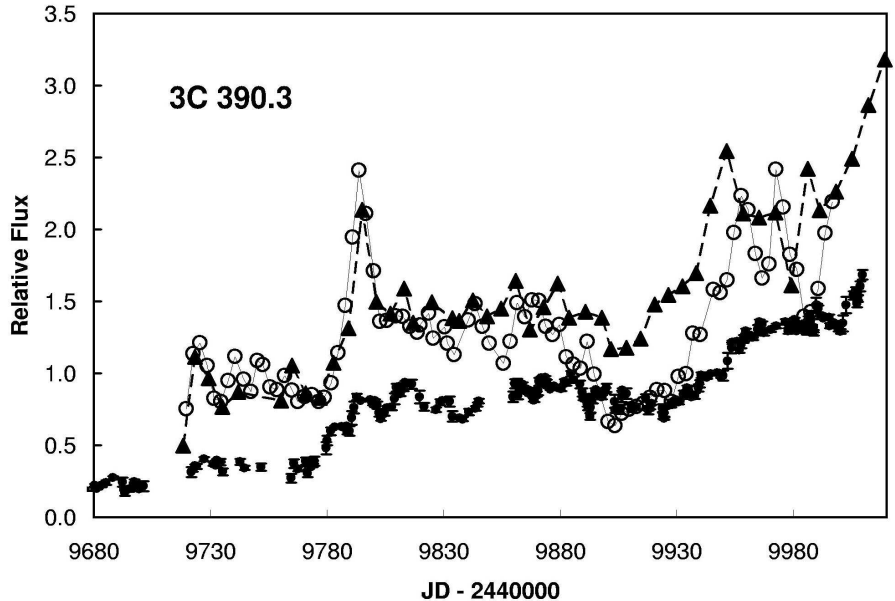} \caption{X-ray, UV,
and optical fluxes of 3C 390.3.  The open circles are 0.1 -- 2 keV
{\it ROSAT} HRI fluxes; the black triangles are {\it IUE}
$\lambda$1370 UV fluxes, and the small black circles with error bars
are V-band fluxes. Scale factors have been chosen for plotting
convenience, and an arbitrary constant has been subtracted from the
V-band fluxes.\label{3C390fast}}
\end{center}
\end{figure}

Fig.~3 shows a number of interesting things.  The strong soft X-ray
flare at JD 2449795 is followed only two days later by a
$\lambda$1370 UV flare that closely matches it in amplitude and
duration, yet in the optical almost nothing happens! The same seems
to be true for a smaller flare at JD 2449724.  However, a pair of
similarly strong soft X-ray flares at JDs 2449957 and 2449972 not
only lack optical counterparts, but also lack UV counterparts.

The long-term variability (see Fig.~4) shows additional interesting
things.  Here we give 24-day average fluxes.  There is now more
qualitative agreement in the shapes and relative amplitudes of the
light curves.  The smoothed optical curves seems to be always
lagging the UV by about three weeks.  Before JD 2449830 the UV seems
to be closely following the soft X-rays.  However, from JD 2449900
onwards a remarkable thing is happening: the UV is clearly {\it
leading} the soft X-rays by $\sim $ 1 week!  The optical is also
leading the X-rays (although probably lagging the UV).  Looking at
the shorter-term variability in Fig. 3, it can be seen that the
increased activity that leads to the strong X-ray flare at JD
2449957 {\it begins in the UV}.

\begin{figure}[t!]
\begin{center}
\epsfxsize = 120 mm \epsfbox{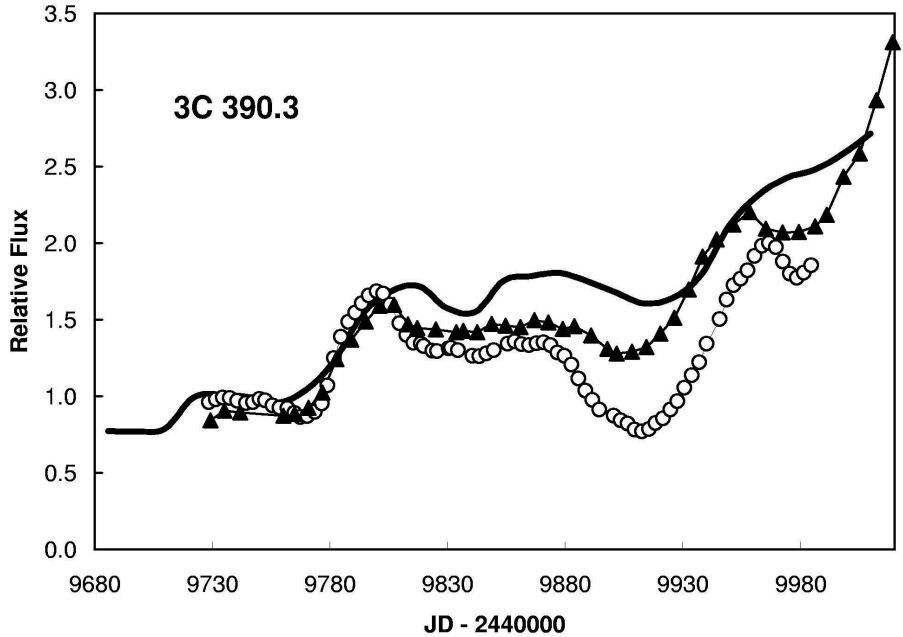} \caption{24-day
running average X-ray, UV, and optical fluxes for 3C 390.3. Open
circles are 0.1 -- 2 keV {\it ROSAT} HRI fluxes; black triangles are
{\it IUE} $\lambda$1370 UV fluxes, and the solid line is the V-band
flux. The scale factors of the X-rays and UV are as in Fig. 3, but
the optical observations have been arbitrarily re-scaled to roughly
match the ranges of the X-ray and UV light curves.
\label{3C390slow}}
\end{center}
\end{figure}

On the other hand, in subsequent monitoring of 3C~390.3 we find that
our V-band flux appears to lag the {\it RXTE} 2--10 keV flux by
$\sim 4.5$ days (see Fig. 4 of \citealt{GaskellKlimek03}), and the
amplitude of the short-term optical variability exceeds the
amplitude of the X-ray variability.

\section{Things to be Explained}

The picture that emerges from these three objects (and others) is a
complicated one.  The following general statements can be made:

(i) \,\,\,\,X-ray variability can be very rapid.

(ii) \,\,\,Optical variability can sometimes be rapid too.

(iii) \,For the very different AGNs NGC~4051, NGC~4151, NGC~5548,
and 3C~390.3, the general level of activity in both the optical and
X-ray regions is correlated on {\it long} timescales, and is
probably of similar amplitude.

(iv) \,\,Short-term variations of the X-ray and optical fluxes {\it
can} be correlated.

(v) \,\,\,\,Short-term variations in the X-ray and optical can also
be {\it un}correlated.

(vi) \,When the optical and X-ray fluxes are correlated, the lag of
the optical is small.

(vii) There is probably a preference for the EUV region to lead
variability, but different signs of phase differences can be seen in
the same object at different times.

\section{An Anisotropic Emission Model}

The frequent independence of the X-ray and optical variability on
short timescales is puzzling given that sometimes the short-term
variability {\it does} appear to be correlated. I want to conclude
by suggesting a possible explanation of this.  I believe that the
rapid soft X-ray and EUV variability implies that the origin of the
variability is {\it electromagnetic}, rather than arising primarily
from gas dynamics. Electromagnetic phenomena propagate close to the
speed of light, while gas-dynamical timescales (which tend to be of
the order of the sound-crossing timescale) are much too long. It has
already been suggested that the large amplitudes observed on short
times in some cases could even require relativistic beaming
\citep[e.g.,][]{Boller97,Reeves02}. An electromagnetic origin of
variability is not unreasonable because, if one believes in energy
equipartition, $E_{magnetic} \sim E_{kinetic}$.

What I want to draw attention to is that electromagnetic variability
{\it is intrinsically anisotropic} -- we see this in the outer
layers of the sun all the time.  This is true even if we are not
seeing relativistic beaming, because particles are accelerated
anisotropically.  The intrinsic anisotropy implies that (a) we will
not always have the emission aimed towards us, and (b) the emitted
electromagnetic radiation or high-energy particles will not always
hit a reprocessing medium. These two considerations result in the
three possibilities that are sketched in Fig. 5 and described in the
caption.  Note that the heating is not necessarily arising from
anisotropic X-ray or EUV emission, but, as with solar flares, it
could arise from high-energy particles hitting the reprocessing
medium.

\begin{figure}[t!]
\begin{center}
\epsfxsize = 100 mm \epsfbox{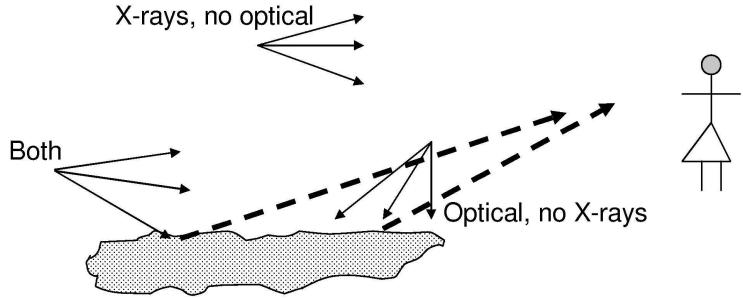} \caption{A schematic
illustration of possible effects of directional emission of X-rays
and high-energy particles on short-term X-ray and optical
variability. The X-ray emission at the top happens to be towards the
observer but mostly misses the reprocessing surface. The emission on
the left is aimed towards the observer, but also hits a reprocessing
surface and produces enhanced optical emission. The X-rays on the
right are not directed at the observer, but do hit the reprocessing
surface. The observer only sees the reprocessed radiation in this
case. \label{model}}
\end{center}
\end{figure}

This model offers an explanation of why short-term X-ray and optical
variability are only sometimes correlated. The small X-ray optical
lags that are seen are on the light-travel time + heating/cooling
timescale. Because optical emission is more isotropic we see more
optical events, but the larger number, plus the smearing out of
optical events, means that the optical amplitude will be smaller.

\section{The Long-Term Relationship Between Optical and
X-ray Variability}

Although reprocessing of soft X-rays and EUV radiation into
longer-wavelength emission fits in with the model just suggested to
explain the varied short-term relationships between X-ray and
optical variability, I think that reprocessing can be ruled out as
the cause of {\it long-term} optical variability for at least three
reasons: first, as shown above, UV and optical variability sometimes
leads X-ray variability. Second, as noted by \citet{Uttley03}, in
NGC~5548 the amplitude of optical variability exceeds that of the
X-ray variability. A third reason is the lack of lag between the
smoothed light curves.  This means that the smoothing function
extends to both positive and negative time, and hence that the
activity in each wavelength region is correlated with the activity
in the other wavelength region at a {\it later} time.  To avoid
violating causality the activity in the separate wavelength regions
must both be driven by some independent slowly varying process.  It
has been shown elsewhere \citet{Gaskell04} that rapid variability
depending on a slowly varying level of activity explains the
proportionality of the variability to the mean flux level, explains
the log-normal nature of X-ray variability, and produces light
curves similar to those observed.

\begin{acknowledgments}
This research is supported by the National Science Foundation
through grant AST 03-07912.
\end{acknowledgments}

\end{document}